\documentclass[prl,letterpaper,twocolumn,showpacs,superscriptaddress,byrevtex]{revtex4}

\usepackage{graphicx,amssymb,amsmath,setspace,stmaryrd}

\begin{document}

\title{Exactly Solvable Model of Monomer-Monomer  Reactions \\
on a  Two-Dimensional Random Catalytic Substrate}

\author{G. Oshanin}
\email{oshanin@lptl.jussieu.fr}
\affiliation{Laboratoire de Physique Th{\'e}orique des Liquides,
Universit{\'e} Paris 6, 4 Place Jussieu, 75252 Paris, France}
\affiliation{
Max-Planck-Institut f\"ur Metallforschung, Heisenbergstr. 3,
D-70569 Stuttgart, Germany}
\affiliation{Institut f\"ur Theoretische und Angewandte Physik,
Universit\"at Stuttgart, Pfaffenwaldring 57, D-70569 Stuttgart,
Germany}
\author{M. N. Popescu}
\email{popescu@mf.mpg.de}
\affiliation{
Max-Planck-Institut f\"ur Metallforschung, Heisenbergstr. 3,
D-70569 Stuttgart, Germany}
\affiliation{Institut f\"ur Theoretische und Angewandte Physik,
Universit\"at Stuttgart, Pfaffenwaldring 57, D-70569 Stuttgart,
Germany}
\author{S. Dietrich}
\email{dietrich@mf.mpg.de}
\affiliation{
Max-Planck-Institut f\"ur Metallforschung, Heisenbergstr. 3,
D-70569 Stuttgart, Germany}
\affiliation{Institut f\"ur Theoretische und Angewandte Physik,
Universit\"at Stuttgart, Pfaffenwaldring 57, D-70569 Stuttgart,
Germany}

\begin{abstract}
We study the equilibrium properties of a monomer-monomer
$A + B \to \emptyset$ reaction on a two-dimensional substrate containing
randomly placed catalytic bonds. Interacting $A$ and $B$ species undergo
continuous exchanges with particle reservoirs and react instantaneously
as soon as a pair of unlike particles is connected by a catalytic bond.
For the case of \textit{annealed} disorder in the placement of catalytic
bonds the model is mapped onto a general spin $S = 1$ model and solved
exactly for the pressure in a particular case. At equal activities of the
two species a second order phase transition is revealed.
\end{abstract}


\maketitle

Catalytically activated reactions (CARs) involve particles which react
only in the presence of another agent -- a catalyst, and remain chemically
inactive otherwise. These processes are widespread in nature and used in a
variety of technological and industrial applications \cite{1b}.

Following
the seminal work of Ziff, Gulari, and Barshad (ZGB) \cite{zgb} on the
so-called ``monomer-dimer'' model, as well as subsequent studies of a simpler
``monomer-monomer'' reaction model \cite{fich,red}, there has been
considerable progress in the understanding of CARs properties. The ZGB model,
which describes, in particular, the important process of ${\rm CO}$ oxidation
on a catalytic surface, revealed several remarkable features \cite{zgb}.
On a two-dimensional (2D) substrate, upon lowering the ${\rm CO}$
adsorption rate the system undergoes a first-order phase transition from a
${\rm CO}$ saturated inactive phase into a reactive steady-state, followed by a
continuous transition into an ${\rm O_2}$-saturated inactive phase, which belongs
to the same universality class as directed percolation and the Reggeon field
theory \cite{universality_class}.
The monomer-monomer model exhibits a first-order transition from a phase saturated
with one species to one saturated with the other; allowing desorption of one
species leads to a continuous transition which also belongs to the directed
percolation universality class \cite{red}.
For these two models, different aspects of the dynamics of the adsorbed phase
have been investigated \cite{zgb,fich,red,con,universality_class,dic}, confirming
an essentially collective behavior. In contrast, the equilibrium properties of
CARs are much less studied and the understanding of the equilibrium state remains
rather limited. Recently, an exact solution for a one-dimensional
monomer-monomer model has been presented \cite{pop}, but for the physically
important 2D substrates no exact solutions are known as yet.

Although realistic substrates are typically disordered and the actual catalyst
is an assembly of mobile or localized catalytic sites or islands \cite{1b},
only a few studies have addressed the behavior of CARs on disordered substrates
focusing on adsorption/desorption processes but not on the coverage of a catalyst
decorated substrate \cite{dis}.
With the exception of few exactly solvable 1D models of $A + A \to \emptyset$
reactions \cite{osh1} and a Smoluchowski-type analysis of $d$-dimensional
CARs \cite{osh2}, for random spatial distributions of the catalyst only empirical
approaches based on phenomenological generalizations of the mean-field ``law of mass
action'' have been proposed so far \cite{1b}. Consequently, an exact analytical
solution of (albeit idealized) models involving a 2D random catalytic substrate is
very desirable since it provides valuable insight into the effects of disorder on the
CARs properties.

In the following we present such an \textit{exactly solvable} model of a monomer-monomer
$A + B \to \emptyset$ reaction on a 2D inhomogeneous, catalytic substrate and study
the equilibrium properties of the two-species adsorbate. The substrate contains
randomly placed catalytic bonds of mean density $q$ which connect neighboring
adsorption sites. The interacting $A$ and $B$ (monomer) species undergo continuous
exchanges with corresponding adjacent gaseous reservoirs. A reaction
$A + B \to \emptyset$ takes place instantaneously if $A$ and $B$ particles occupy
adsorption sites connected by a catalytic bond. We find that for the case of
\textit{annealed} disorder in the placement of the catalytic bonds the reaction model
under study can be mapped onto the general spin $S = 1$ (GS1) model \cite{berker}.
This allows us to exploit the large number of results obtained for the GS1 model
\cite{berker} in order to elucidate the equilibrium properties of the monomer-monomer
reaction on random catalytic substrates \cite{pop2}. Here we concentrate on a
particular case in which the model reduces to an exactly solvable
Blume-Emery-Griffiths (BEG) model \cite{BEG,har} and derive an exact expression for
the disorder-averaged equilibrium pressure of the two-species adsorbate. We show that
at equal partial vapor pressures of the $A$ and $B$ species this system exhibits a
second-order phase transition which reflects a spontaneous symmetry breaking with
large fluctuations and progressive coverage of the entire substrate by either one
of the species.

\begin{figure}[!htb]
\begin{center}
\includegraphics[width=.55 \linewidth]{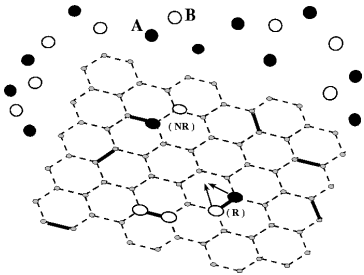}
\caption{
\label{fig1}
2D lattice of adsorption sites (small grey circles) in contact with a mixed vapor
phase. Black and white circles denote $A$ and $B$ particles, respectively. The solid
lines denote ``bonds''  with catalytic properties. (\textbf{R}): configuration in
which an instantaneous reaction takes place ({\tiny $\nearrow \nwarrow$}) upon which
the reactants leave the system. (\textbf{NR}): NN pair of $A$ and $B$ which do not
react since the sites are not connected by a catalytic bond.
}
\end{center}
\end{figure}

We consider a 2D regular lattice of $N$ adsorption sites (Fig.~\ref{fig1}), which
is in contact with the mixed vapor phase of $A$ and $B$ particles.
The $A$ and $B$ particles can adsorb onto $vacant$ sites, and can desorb back to
the reservoir. They are characterized by chemical potentials $\mu_A$ and $\mu_B$
which are maintained at constant values and measured relative to the binding energy
of an occupied site, so that $\mu_{A,B} > 0$ corresponds to a preference for
adsorption. The $A$ and $B$ particles have hard cores prohibiting double occupancy
of sites and nearest-neighbor (NN) attractive $A-A$, $B-B$, and $A-B$ interactions
of strengths $J_A$, $J_B$, and $J_{AB}$, respectively. The occupation of the $i$-th
site is described by a pair $c_i$ of Boolean variables $n_i$ and $m_i$ such that
\begin{equation}
\label{variables}
c_i \equiv (n_i,m_i) =
\begin{cases}
(1,0), & \textrm{site $i$ occupied by $A$,} \cr
(0,1), & \textrm{site $i$ occupied by $B$,} \cr
(0,0), & \textrm{site $i$ empty,} \cr
(1,1), & \textrm{excluded by hard cores.} \cr
\end{cases}
\end{equation}

We assign to some of the lattice bonds (solid lines in Fig.~\ref{fig1}) ``catalytic''
properties such that if an $A$ and a $B$ particle occupy simultaneously NN sites
connected by such a catalytic bond, they instantaneously $react$ by forming a product
($\emptyset$) which immediately desorbs and leaves the system; $A$ and $B$ particles
occupying NN sites not connected by a catalytic bond harmlessly coexist.
The ``catalytic'' character of the lattice bonds is described by Boolean variables
$\zeta_{<ij>}$, where $<ij>$ denotes a pair of neighboring sites $i$ and $j$,
\begin{equation}
\zeta_{<ij>} =
\begin{cases}
1, & ~~\textrm{$<ij>$~is a catalytic bond,} \cr
0,  & ~~\textrm{otherwise,}\cr
\end{cases}
\label{zeta}
\end{equation}
and we take $\{\zeta_{<ij>}\}$ as independent, identically distributed random variables
with the distribution
\begin{equation}
\varrho(\zeta) =  q \delta(\zeta - 1) + (1-q)\delta(\zeta).
\label{dist_zeta}
\end{equation}
The probability $q$ that a given bond is catalytic equals the mean density of the
catalytic bonds. The two limiting cases, $q = 0$ and $q = 1$, correspond to an
\textit{inert} substrate and to a \textit{homogeneous catalytic} one, respectively.

The condition of instantaneous reaction $A+B \to \emptyset$ is formally equivalent to
allowing a NN $A-B$ repulsive interaction of strength $\lambda \to \infty$ for $A-B$
pairs connected by catalytic bonds. Hence, in thermal equilibrium and for a given
configuration $\zeta \equiv \{\zeta_{<ij>}\}$, the partition function of such a
two-species adsorbate is
\begin{equation}
\label{partition}
Z_N(\zeta) = \lim_{\lambda \to \infty}\sum_{\{c_k\}}
\exp\left[-\beta {\cal H}_\lambda(\zeta)\right],
\end{equation}
where $\beta^{-1} = k_B T$ is the thermal energy, while the Hamiltonian
${\cal H}_\lambda(\zeta) = H_\lambda(\zeta) + H_0$ naturally separates into
a disorder-dependent part,
\begin{equation}
\label{disorder_H}
H_\lambda(\zeta) = \lambda \sum_{\langle ij \rangle}
\zeta_{<ij>} \left( n_i m_j + n_j m_i \right),
\end{equation}
where the summation extends over all pairs $\langle ij \rangle$,
and a disorder-independent contribution
\begin{eqnarray}
\label{regular_H}
H_0 &=& -\sum_{\langle ij \rangle}
\left[J_{A} n_i n_j + J_{B} m_i m_j
+ J_{AB}\left( n_i m_j + n_j m_i \right) \right]\nonumber\\
&&-\sum_{i=1}^{N}\left(\mu_A n_i+\mu_B m_i\right).
\end{eqnarray}

In what  follows we shall focus on situations in which the  disorder in the
placement of the catalytic bonds is \textit{annealed}. In this case the
thermodynamics of the system is given by the disorder-averaged pressure (in units of
the lattice cell area),
\begin{equation}
P \equiv P(T,\mu_A,\mu_B) = \frac{1}{\beta}
\lim_{N \to \infty} \frac{1}{N} \ln \langle Z_N(\zeta)\rangle_\zeta \,,
\label{pressure}
\end{equation}
where $\langle \dots \rangle_\zeta$ denotes the average over all possible
realizations $\{\zeta_{<ij>}\}$. Once $P$ is known all other thermodynamic
quantities of interest can be obtained by differentiating $P$ with respect to
$\mu_A$, $\mu_B$, or $T$.

Averaging $Z_N(\zeta)$ in Eq.~(\ref{partition}) is straightforward,
\begin{eqnarray}
&&\langle Z_N(\zeta)\rangle_\zeta = \sum_{\{c_k\}} e^{-\beta H_0}
\lim_{\lambda \to \infty}
\langle e^{-\beta H_\lambda(\zeta)}\rangle_\zeta
\nonumber\\
&=& \sum_{\{c_k\}} e^{-\beta H_0} \prod_{\langle ij \rangle}
\lim_{\lambda \to \infty}
\left[q \, e^{- \lambda \beta \left( n_i m_j + n_j m_i \right)} + 1-q \right]
\nonumber\\
&=& \sum_{\{c_k\}} e^{-\beta H_0}
\prod_{\langle ij \rangle} (1-q)^{n_i m_j + n_j m_i}
=\sum_{\{c_k\}} e^{-\beta {\cal H}_{e}}\,\,,
\label{factor_1}
\end{eqnarray}
and yields the "effective" Hamiltonian
\begin{eqnarray}
\label{Heff}
{\cal H}_{e} &=&
-\sum_{\langle ij \rangle} \left\{
\left[J_{AB}+\beta^{-1} \ln(1-q)\right] \left(n_i m_j + n_j m_i \right)
\right.
\nonumber\\
&+&\left.
J_{A} n_i n_j+ J_{B} m_i m_j \right\}-\sum_{i=1}^{N}\left(\mu_A n_i+\mu_B m_i\right).
\end{eqnarray}
Introducing the ``spin'' variables $\sigma_i \in \{0,\pm 1\}$,
\begin{equation}
\sigma_i =
\begin{cases}
+(-)1, & ~~\textrm{site $i$ occupied by $A$ ($B$),} \cr
~0,  & ~~\textrm{site $i$ empty,}\cr
\end{cases}
\label{spin}
\end{equation}
such that $n_i=(\sigma_i+\sigma_i^2)/2$ and $m_i=(-\sigma_i+\sigma_i^2)/2$,
${\cal H}_{e}$ can be cast into the form of the classical Hamiltonian of the
general spin $S = 1$ model \cite{berker},
\begin{eqnarray}
&&\hspace*{.3in}\mathcal{H}_{e} = - J\sum_{<ij>}\sigma_i \sigma_j -
K \sum_{<ij>}\sigma_i^2 \sigma_j^2 \nonumber\\
&&-C\sum_{<ij>}\left(\sigma_i \sigma_j^2+\sigma_j \sigma_i^2\right)
-H \sum_{i=1}^{N} \sigma_i + \Delta \sum_{i=1}^{N} \sigma_i^2\hspace*{.3in}
\label{H_BEG}
\end{eqnarray}
with coupling constants
\begin{subequations}
\label{param}
\begin{eqnarray}
&&J = \frac{J_A+J_B-2 J_{AB}}{4} -\frac{\ln(1-q)}{2 \beta}, \hspace*{.3in}
\label{J_par}\\
&&K = \frac{J_A+J_B+2 J_{AB}}{4} + \frac{\ln(1-q)}{2 \beta}, \hspace*{.3in}
\label{K_par}\\
&&C = \frac{J_A-J_B}{4},\; H = \frac{\mu_A -\mu_B}{2},\;
\Delta = -\frac{\mu_A +\mu_B}{2}.\hspace*{.4in}
\label{D_par}
\end{eqnarray}
\end{subequations}
Thus, in the case of annealed disorder, the $A + B \to \emptyset$ reaction model
under study can be mapped exactly onto the GS1 model, which has been extensively
analyzed \cite{berker}. The accumulated knowledge
on its critical behavior, phase diagrams, as well as low- and high-temperature
expansions \cite{berker} can be straightforwardly used to elucidate the
equilibrium properties of the present CAR model for general values of $\mu_A$,
$\mu_B$, $J_A$, $J_B$, $J_{AB}$, and $q$, as well as for different types of
embedding lattices \cite{pop2}.

In the remaining part of this paper we focus on the symmetric case
$\mu_A = \mu_B$ and $J_A = J_B$, implying $C = H = 0$ so that the model reduces
to the original Blume-Emery-Griffiths (BEG) model \cite{BEG}. Additionally, we
set $J_{AB} = 0$ and consider a honeycomb lattice and a particular relation
between $K$ and $J$, for which the 2D BEG model, and hence the monomer-monomer
reaction model under study, can be solved \textit{exactly} \cite{har,note}.

Following Ref.~\cite{har}, in the subspace
\begin{equation}
e^{-\beta K} = \cosh(\beta J),
\label{subspace}
\end{equation}
the partition function of the 2D BEG model on the honeycomb lattice may be
expressed in terms of the partition function of a zero-field Ising model on the
honeycomb lattice, which is known in closed form \cite{Lavis}. In this subspace
the 2D BEG model exhibits an Ising type phase transition with a line of critical
points obeying \cite{har}
\begin{equation}
\tanh(\beta J) = \frac{2 + e^{\beta \Delta}}{2 \sqrt{3}}.
\label{T_c}
\end{equation}
For our CAR model Eq.~(\ref{subspace}) implies
\begin{equation}
\beta J_A = \beta J_B = \ln\left(1 + q \right).
\label{log}
\end{equation}
In the subspace defined by Eq.~(\ref{log}) the disorder-averaged pressure
(Eq.~(\ref{pressure})) is given \textit{exactly} by
\begin{equation}
\label{pressure2}
\beta P = \frac{1}{2}\left(1 + 2 e^{\beta \mu}\right) -
\frac{3}{2} \cosh(\beta{\cal J}) - f({\cal J}),
\end{equation}
where $\mu = \mu_A = \mu_B$, $ \beta {\cal J} =
\tanh^{-1}\left(2 q/(2+e^{-\beta \mu})\right)$,
and $f({\cal J})$ denotes the known free energy of a spin-$1/2$ Ising model with NN
interaction ${\cal J}$ on a honeycomb lattice (see Ref. \cite{Lavis}). Since in this
particular case the exact expression for the average ``magnetization''
$M_0 =  \langle \sigma_i \rangle$ is also known \cite{Urumov}, the average densities
of $A$ and $B$ species are straightforwardly calculated as
$\langle n_i \rangle = (M_0+\langle \sigma_i^2 \rangle)/2$ and
$\langle m_i \rangle=(-M_0+\langle \sigma_i^2 \rangle)/2$, where
$\langle \sigma_i^2 \rangle = z \partial_z (\beta P)$ and $z = e^{\beta \mu}$.
Furthermore, the line of critical points as function of $q$ (within the subspace
defined by Eq.~(\ref{log})) where a continuous transition takes place is given by
\begin{equation}
\beta \mu_c = - \ln[2 \, (q \sqrt{3}-1)].
\label{T_cc}
\end{equation}

\begin{figure}[!htb]
\begin{minipage}[c]{.89\linewidth}
\includegraphics[width=.95\linewidth]{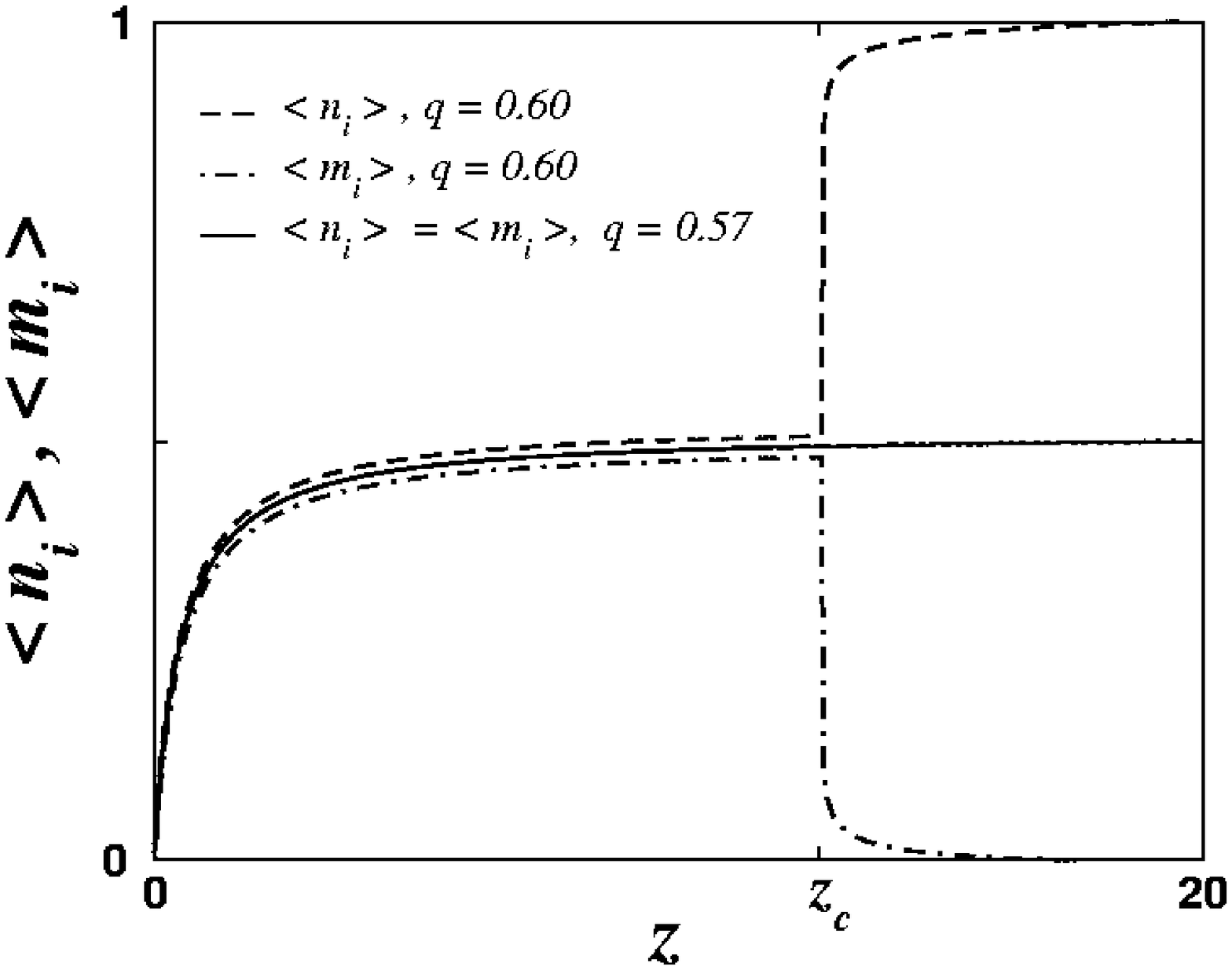}%
\raisebox{.1\linewidth}
{
\hspace{-.7\linewidth}%
\includegraphics[width=.3\linewidth]{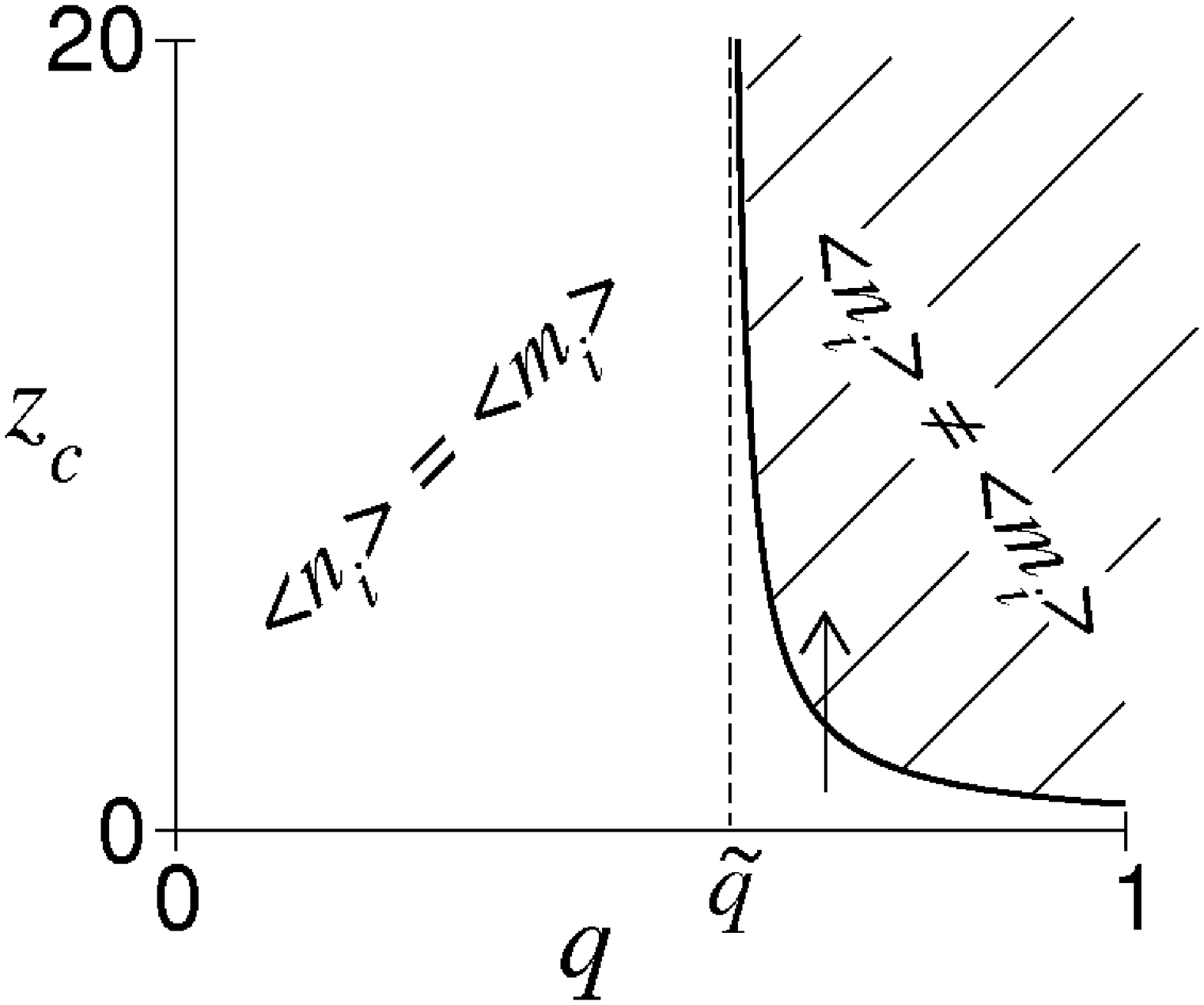}%
}%
\raisebox{.435\linewidth}
{
\hspace{.09\linewidth}%
\includegraphics[width=.25\linewidth]{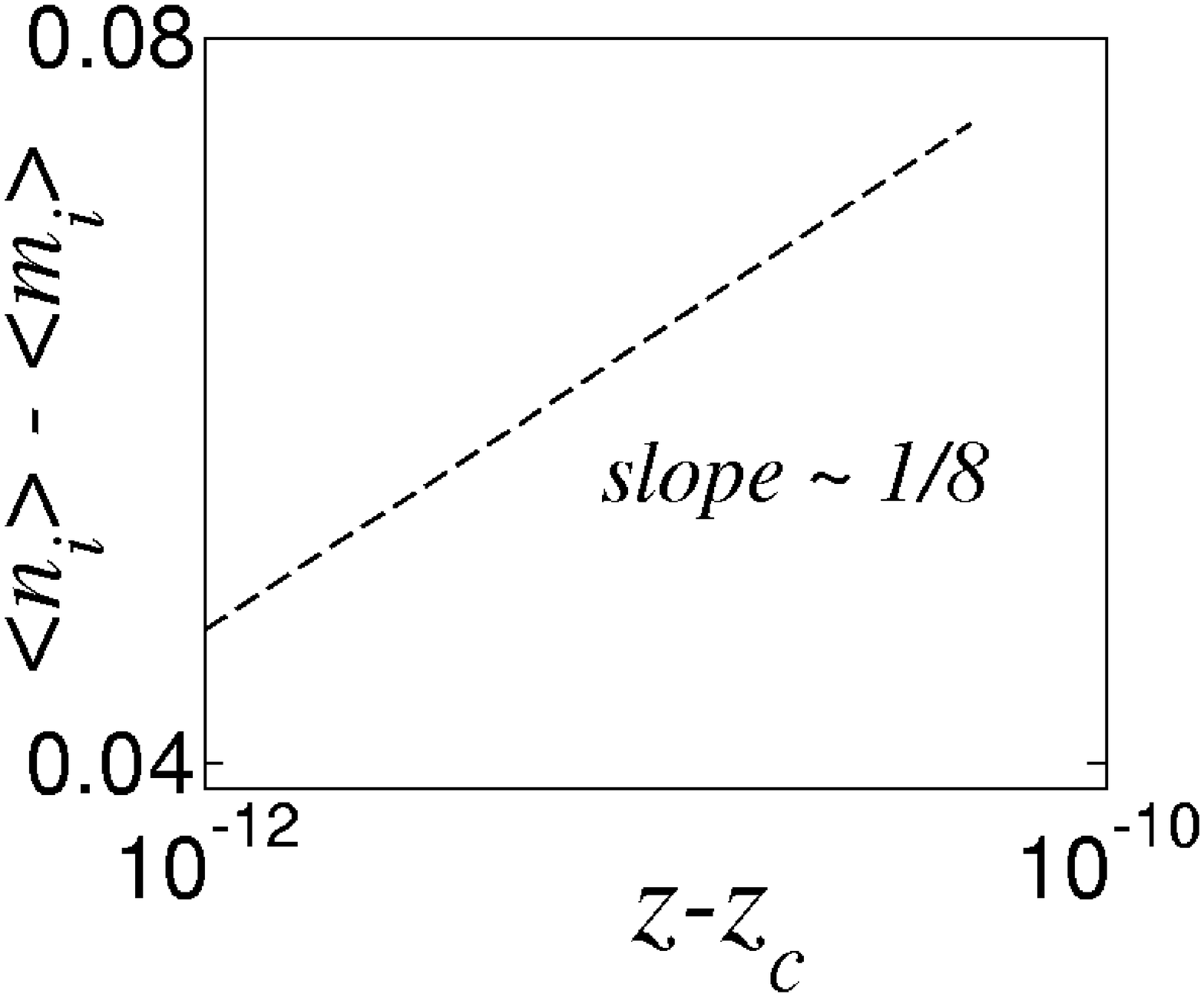}%
}
\end{minipage}
\caption{
\label{fig2}
Average density of $A$ particles, $\langle n_i \rangle$, and of $B$ particles,
$\langle m_i \rangle$, as function of their common fugacity $z$ above ($q = 0.60$)
and below ($q = 0.57$) the threshold value $\tilde q = 1/\sqrt{3}$ for the
concentration of catalytic segments. For clarity, the curves corresponding to
$q = 0.60$ have been symmetrically shifted up ($\langle n_i \rangle$) and down
($\langle m_i \rangle$), respectively. The upper inset (logarithmic scale)
illustrates the scaling behavior $\langle n_i \rangle-\langle m_i \rangle
\sim (z-z_c)^{1/8}$ for $z \to z_c$. The lower inset shows the critical line
$z_c(q)$ (Eq.~(\ref{T_cc})); the arrow indicates a path of increasing $z$
at fixed $q$ which crosses the transition line.
}
\end{figure}

We now emphasize several features of these results. {\bf(i)} For $\mu$ below its
critical value (Eq.~(\ref{T_cc})) we find
$\langle n_i \rangle = \langle m_i \rangle \geq 0$ (see also Fig.~\ref{fig2}).
Upon exceeding $\mu_c$ (by increasing the vapor pressure in the reservoirs)
one of the densities (with equal probability) decreases sharply but
\textit{continuously} to zero while the other one rapidly attains unity.
This reveals a spontaneous symmetry breaking and implies that the substrate
becomes poisoned, i.e., most of it is covered by either one of the species.
If the chemical potentials $\mu_A$ and $\mu_B$ differ slightly, the transition
to the poisoned state is smeared out but remains detectable.
{\bf(ii)} The transition can occur only if the mean density $q$ of catalytic bonds
$q$ is sufficiently high, such that $q > \tilde{q}=1/\sqrt{3} \approx 0.577$.
For $q < \tilde{q}$, $\langle n_i \rangle = \langle m_i \rangle$ for all
$\beta \mu$, and both tend to $1/2$ as $\beta \mu \to \infty$.
{\bf(iii)} $\mu_c \geq 0$ for $q \in \left[\frac{1}{\sqrt{3}},\frac{\sqrt{3}}{2}\right]$,
which means that in this range of $q$ values the transition occurs in situations
in which adsorption on the substrate is favored. For
$q \in \left(\frac{\sqrt{3}}{2},1 \right]$, $\mu_c < 0$ and hence the transition
takes place for the case that desorption into the reservoir is favored.
{\bf(iv)} There occur large scale critical fluctuations of the densities of adsorbed $A$
and $B$ particles upon approaching $z_c = \exp(\beta \mu_c)$ from above or
below (by varying the vapor pressure in the reservoirs), and the compressibility of
the adsorbed phase \textit{diverges} as $|z-z_c|^{-7/4}$ for $z \to z_c$ and
$\mu_A = \mu_B$, and as $|\mu_A-\mu_B|^{-14/15}$ for $\mu_A \to \mu_B$
and $z = z_c$ \cite{Pelissetto}.
{\bf(v)} An analysis of the model on a Bethe lattice (coordination number $\gamma = 3$)
shows that the case $e^{\beta K} \cosh(\beta J) = 1$ is not singular, i.e.,
the transition line discussed above persists for $e^{\beta K} \cosh(\beta J) \lesssim 1$
and for $1< e^{\beta K} \cosh(\beta J) <2$ \cite{pop2}.
We finally remark
that one should
not expect big differences in the critical behavior in the case of \textit{quenched}
disorder because the transition occurs only for sufficiently large values of
$q$ \cite{osh1}.

In conclusion, this study presents an \textit{exactly solvable} model of a
monomer-monomer $A + B \to \emptyset$ reaction on a 2D random catalytic substrate.
This exact solution has been obtained via a mapping of the partition function of
the two-species adsorbate onto the partition function of a general spin $S=1$
model and by noticing that for certain relations between the corresponding
coupling constants the latter reduces to an exactly solvable 2D BEG
model \cite{har}. In this case we have determined the annealed disorder-averaged
equilibrium pressure of the two-species adsorbate and have shown that the system
under study exhibits a second-order (-- robust in parameter space --) 2D Ising-like
phase transition if the mean density of the catalyst is sufficiently large.

GO thanks the AvH Foundation for financial support via the Bessel
Forschungspreis. MNP acknowledges very instructive discussions with K. Mecke
and D. Dantchev.

\end{document}